%Paper: hep-ph/9207204
%From: Jean Cleymans <UPHYF584@comparex.hrz.uni-bielefeld.de>
%Date: Thu, 02 Jul 92 11:57:28 CET

\magnification=1200
\hsize=16.0 truecm
\vsize=23.0 truecm
\baselineskip=13 pt
\footline={\ifnum\pageno=0\hfil\else\hss\tenrm\folio\hss\fi}
\pageno=0
\def\J{$J/\psi$}
\def\T{$T_f$}
\def\gsim{\raise0.3ex\hbox{$>$\kern-0.75em\raise-1.1ex\hbox{$\sim$}}}
\newcount\REFERENCENUMBER\REFERENCENUMBER=0
\def\reftag#1{\expandafter\ifx\csname RF#1\endcsname\relax
               \global\advance\REFERENCENUMBER by 1
               \expandafter\xdef\csname RF#1\endcsname
                      {\the\REFERENCENUMBER}\fi
             \csname RF#1\endcsname\relax}
\def\refto#1#2{\expandafter\ifx\csname RF#1\endcsname\relax
               \global\advance\REFERENCENUMBER by 1
               \expandafter\xdef\csname RF#1\endcsname
                      {\the\REFERENCENUMBER}\fi
           \expandafter\ifx\csname RF#2\endcsname\relax
               \global\advance\REFERENCENUMBER by 1
               \expandafter\xdef\csname RF#2\endcsname
                      {\the\REFERENCENUMBER}\fi
             [\csname RF#1\endcsname--\csname RF#2\endcsname]\relax}
\def\ref#1{\expandafter\ifx\csname RF#1\endcsname\relax
               \global\advance\REFERENCENUMBER by 1
               \expandafter\xdef\csname RF#1\endcsname
                      {\the\REFERENCENUMBER}\fi
             [\csname RF#1\endcsname]\relax}
\def\refs#1#2{\expandafter\ifx\csname RF#1\endcsname\relax
               \global\advance\REFERENCENUMBER by 1
               \expandafter\xdef\csname RF#1\endcsname
                      {\the\REFERENCENUMBER}\fi
           \expandafter\ifx\csname RF#2\endcsname\relax
               \global\advance\REFERENCENUMBER by 1
               \expandafter\xdef\csname RF#2\endcsname
                      {\the\REFERENCENUMBER}\fi
            [\csname RF#1\endcsname,\csname RF#2\endcsname]\relax}
\def\refss#1#2#3{\expandafter\ifx\csname RF#1\endcsname\relax
               \global\advance\REFERENCENUMBER by 1
               \expandafter\xdef\csname RF#1\endcsname
                      {\the\REFERENCENUMBER}\fi
           \expandafter\ifx\csname RF#2\endcsname\relax
               \global\advance\REFERENCENUMBER by 1
               \expandafter\xdef\csname RF#2\endcsname
                      {\the\REFERENCENUMBER}\fi
           \expandafter\ifx\csname RF#3\endcsname\relax
               \global\advance\REFERENCENUMBER by 1
               \expandafter\xdef\csname RF#3\endcsname
                      {\the\REFERENCENUMBER}\fi
[\csname RF#1\endcsname,\csname RF#2\endcsname,\csname
RF#3\endcsname]\relax}
\def\refadd#1{\expandafter\ifx\csname RF#1\endcsname\relax
               \global\advance\REFERENCENUMBER by 1
               \expandafter\xdef\csname RF#1\endcsname
                      {\the\REFERENCENUMBER}\fi \relax}

%
%\hfill 30.6.1992\par
\hfill CERN-TH.6523/92\par
\hfill BI-TP 92/08 \par
\vskip 2 truecm
\font\titlefont=cmbx10 scaled \magstep1
\begingroup\titlefont\obeylines
\centerline{THERMAL HADRON PRODUCTION IN}\vskip 5pt
\centerline{HIGH ENERGY HEAVY ION COLLISIONS}
\endgroup\bigskip\bigskip
\parindent=0pt
\centerline{J.~Cleymans$^{1,2}$\footnote{}{1$~~\!$Department of
Physics, University of Cape Town, Rondebosch 7700, South Africa}
\footnote{}{2$~~\!$Fakult\"at f\"ur Physik, Universit\"at Bielefeld,
D-4800 Bielefeld 1, Germany}
and~~H.~Satz$^{2,3}$\footnote{}{3~Theory Division,
CERN, CH-1211 Geneva 23, Switzerland\hfill\break
\vskip 5pt
\noindent CERN-TH.6523/92\hfill\break
\noindent BI-TP 92/08\hfill\break
\noindent May 1992\hfill}}
\vskip 2truecm
\parindent=20pt
\bigskip\centerline{\bf ABSTRACT}\medskip
We provide a method to test if hadrons produced in high energy
heavy ion collisions were emitted at freeze-out from an equilibrium
hadron gas.
Our considerations are based on an ideal gas at fixed temperature $T_f$,
baryon number density $n_B$, and vanishing total strangeness. The
constituents of this gas are all hadron resonances up to a
mass of 2 GeV; they
are taken to decay according to the experimentally observed branching
ratios. The ratios of the various resulting hadron production rates are
tabulated as functions of $T_f$ and $n_B$. These tables can be used for the
equilibration analysis of any heavy ion data; we illustrate this
for some specific cases.
\vfill\eject
\noindent {\bf 1. Introduction}
\bigskip
\refadd{qm90} \refadd{WA85} \refadd{NA35} \refadd{E802} \refadd{NA38}
\refadd{vesa1} \refadd{vesa2} \refadd{amis} \refadd{carsten}
\refadd{strange} \refadd{raf}
Five years ago, an extensive program for the study of strongly
interacting matter
through high energy heavy ion collisions
was initiated at CERN and BNL \refto{qm90}{NA38}
 A crucial element in
this endeavour is the production of systems which are indeed
thermalized -- and not just an uncorrelated superposition of
nucleon-nucleon collisions. Thermalisation can only be achieved
through sufficient rescattering of the incoming nucleons among each
other and with secondaries, and of the secondaries among each other.
It is not obvious if, how and at what stage thermalisation is achieved.
If it is achieved at all, and if it is not destroyed subsequently by
possible non-equilibrium features \refto{vesa1}{raf},
 then the produced hadrons
should be observed in ratios determined by the thermodynamics of a
system at freeze-out temperature $T_f$ and freeze-out baryon density
$n_B$.
The aim of this work is to determine and tabulate these
ratios for the various hadron species as function of $T_f$
and $n_B$. Our
tables should allow a thermalisation analysis of any present and
future heavy ion data, and we shall illustrate how such an analysis
is carried out. We should emphasize that we are not providing a model
of any kind for the space-time development of the systems produced, but
merely a check of the degree of thermalisation indicated in the
observed hadron ratios. In particular, the collisions
between light ion projectiles and heavy ion targets
studied studied up to now very likely do not lead to complete
thermalisation, and so we expect experiments to deviate from the
ratios listed here. These deviations should, however, lead to values
between the ratios from p-p collisions and the thermal ones. If they
don't, this could be a hint for some new effect. Furthermore, if there
is thermalisation, one cannot say anything about how it was
established. Thermal systems have no memory, and hence the observation
of thermal hadron production ratios neither excludes nor confirms
quark deconfinement at an earlier stage.\par
It should be noted that the thermalisation we are considering here
refers to the approach to a statistical equilibrium among the
different hadron species. It is conceivable \ref{raf} that such a
``chemical'' equilibrium sets in later than a ``thermal'' equilibrium
among different hadrons of the same species. So far, however, there
seem to be no clear-cut tests for the latter. \par
\bigskip
\noindent{\bf 2. Thermal Hadron Production}
\bigskip
Hadronic matter in equilibrium is specified by three
parameters : the temperature \T, the
baryon number density $n_B$, and the strangeness density $n_S$;
the latter is zero unless one is interested in globs of strange
matter \ref{strange}.
Hadronic matter in thermal and chemical equilibrium
will have a certain composition of particle species,
characterized by ``chemical'' potentials.
In addition
to the baryon number chemical potential $\mu_B$, we have
a strangeness chemical potential $\mu_S$, since particles of a given
strangeness
can have different baryon numbers (e.g., ${\bar K}^0$ and $\Lambda$).
The value of $\mu_B$ is fixed by giving the overall baryon number
density $n_B$, and that of $\mu_S$ by fixing the overall strangeness,
which in our case will be zero.
The number densities for all hadron states are then given in terms of
only $\mu_B$ and \T. There are many particle species,
and if they are produced in equilibrium, all their ratios must be fixed
in terms of these two parameters. It is thus straightforward to
test if there was
equilibrium, and since this question is so basic, it
should be one of the first tests to be made.\par
We include only the essential
interaction features of a multi-hadron system near freeze-out:
resonance production in general and particle repulsion at high
baryon density. The bulk of meson production
is known to take place through intermediate resonance states, and we
therefore start with an ideal gas containing
the observed (strange and non-strange) mesonic and baryonic resonances.
A crucial question here is how many of the observed resonances we should
include. There is no clear-cut answer -- by including more and more
resonance states, we evidently go back further and further in the
evolution of the system. To obtain a definite scheme, we shall include
all resonances up to a mass of 2 GeV. In order
to test the importance of high
mass contributions, we have also carried out calculations including
only the resonances of the basic multiplets, i.e., the pseudoscalar
meson octet plus singlet, the vector meson octet plus singlet,
the baryon octet and the baryon decuplet. --
To obtain the distributions of the actually observed
hadrons, in particular pions and kaons, we then let the resonances
decay according to the measured branching ratios.
This procedure is well known; see e.g. \ref{heinz}.
\par
For small or
vanishing baryon number, this would suffice to fix our model; at high
$n_B$, however, a strong, short-range repulsion sets in between baryons,
even if they have different quantum numbers;
this can be described in terms of excluded volumes, repulsive interaction
contributions, or Pauli blocking. We shall here adopt
a hard-sphere picture \ref{Redlich}, in which the baryons in the
system effectively remove a portion of the available spatial volume.
 As a consequence, the baryon number density $n_B${} becomes
$$
n_B~={n_B^o\over [1~+~V_0n_B^0]}~~, \eqno(1)
$$
where $n_B^o$ denotes the density calculated for an ideal gas of
pointlike baryons. We note that for $n_B^o\rightarrow\infty$,
$n_B\rightarrow 1/V$, i.e., we have dense
packing\footnote{*}{Actually $1/V$ is the limiting
 value for in\-com\-pres\-si\-ble
but deformable baryons; a true hard-sphere system yields $1/cV$, with
$c\ge 1$ \ref{karsch}.}
of hard-sphere baryons with an intrinsic volume $V$. In the case of
several baryon species $\alpha = 1, 2,...,r$, eq.(1) becomes
$$
n_B~={n_B^o\over [1~+~\sum_{\alpha=1}^r V_{\alpha} n_{\alpha}^0]}~~,
 \eqno(2)
$$
where the baryon volume $V_{\alpha}$ could be proportional to the
mass, as suggested e.g. by the bag model. Here we take one universal
volume
$V=4\pi R^3/3$, with $R=0.8$ fm as an indicative radius for the
nucleon, and apply the excluded volume correction (1) to all particle
species, mesons as well as baryons. Note that this does not imply
repulsion between mesons; it only reduces space accessible to them,
because some is already ``used up" by the hard-core baryons.
For matter at $n_B=0$, the
full volume is once more available; for $T=0$, we recover the
cold nucleon gas with hard-sphere repulsion \ref{Redlich}. The
correction term moreover cancels out in
all particle ratios. -- This completes our picture.
\par
To illustrate the resulting pattern, we
note that
$$
n_{K^{\pm}}^0 = \int {d^3p\over (2\pi)^3}~{1\over [e^{(E_K \mp \mu_S)/T}~-~1]}
\eqno(3)
$$
gives the number density of {\it directly produced thermal} kaons, and
$$
n^0_{\Lambda}~=~\int {d^3p\over (2\pi)^3}~{1\over [e^{(E_{\Lambda}-(\mu_B -
\mu_S))/T}~+~1]} \eqno(4)
$$
that of {\it directly produced thermal, pointlike} $\Lambda$'s;
here $E_K = ({\bf p}^2 + m_K^2)^{1/2}$ and $E_{\Lambda} = ({\bf p}^2
+ m_{\Lambda}^2)^{1/2}$. The actual production rates are obtained
from this by including baryon repulsion according to eq.(2) and by
adding to eqs.(3) and (4) the corresponding values from resonance
production, such as $K^{0*}\rightarrow K^+ + \pi^-$, using the
experimentally determined branching ratios.
\par
Requiring the difference between all strange particle number densities
to vanish,
$$
\sum_i n_S^i~-~\sum_i n_{\bar S}^i ~=~0~, \eqno(5)
$$
fixes $\mu_S$ and thus gives us all particle densities in terms of
$T$ and $\mu_B$. Summing at fixed $\mu_B$ over all baryon
contributions,
$$
\sum_i n^i_B - \sum_i n^i_{\bar B} ~=~n_B~~,\eqno(6)
$$
determines the overall baryon number density $n_B$.  Note that in
eq.(6) we sum over the {\it physical} baryon densities
as defined by eq.(2). Eqs.(5) and (6) provide a relation between the
strange chemical potential and the baryon chemical potential; it is
shown in Fig.1. With increasing temperature, $\mu_S$ is seen to
increase faster with $\mu_B$, and for temperatures $T\gsim$ 200 MeV,
$\mu_S/\mu_B$ is found to be rather constant in the range considered
here. In particular we
note that for $T\simeq 200$ MeV and up to about $\mu_B\simeq $ 500
MeV, we have $\mu_S \simeq (1/3) \mu_S$.
In terms of the baryon number and strangeness chemical
potentials for quarks, $\mu_b$ and $\mu_s$, respectively, the hadronic
chemical potentials become $\mu_B=3\mu_b$ and $\mu_S=\mu_b - \mu_s$.
Hence $\mu_B\simeq 3\mu_S$ just translates into $\mu_s\simeq 0$.
Since this arises here quite naturally in a hadronic picture, it is
difficult to consider it as evidence for a primordial quark-gluon plasma
\ref{Raf-Gat}.
\par
 In Fig.2 we illustrate the effect of hard-core repulsion
between baryons,
by plotting $n_B${} as function of $\mu_B$ for a hard-core
volume $V=0$ and for one with radius $R=0.8$ fm. We note that the
repulsion begins to become crucial around
$n_B${}$\simeq $ 0.1 fm$^{-3}$, i.e., already below standard nuclear
density .
\par
We should emphasize that the particle ratios which we will
discuss in this paper are by construction independent of the specific
relation between the baryon chemical potential
and the baryon density: as already mentioned,
the repulsion introduced in equation (1)
cancels out in the ratios.
The relation between baryon number density and baryon chemical
potential, on the other hand, does depend on the specific form (1).
In the tables given
at the end of this paper, the baryon number density is
determined by this form with the hard core volume as defined there.
\bigskip
\noindent{\bf 3. Thermal Hadron Rates}
\bigskip
In this section we will discuss the general features of the
various hadron species. Since we are here only interested in the
qualitative behaviour, the results for this section were obtained using
only the basic hadron multiplets, unless otherwise noted.
\medskip
\noindent{\bf 3.1 Pions}
\medskip
\refadd{barz} \refadd{kusnezov}
\refadd{brown} \refadd{jsk}
Pions are present in thermal equilibrium in the heat bath and
in addition
in the decay products of almost all heavier particles.
As long as the temperature is below 100 MeV, only very
few heavy particles are present and almost all pions are of direct
thermal origin.
It is well known, however, that the number of directly produced
thermal pions decreases at higher temperatures (see e.g.
\refs{vesa1}{vesa2},\ref{heinz},\refto{barz}{jsk}).
This becomes evident in Fig.3, where the ratio of all pions
(i.e., including decay products) to direct thermal pions is
plotted
as a function of temperature. It increases rapidly above 100 MeV.
For higher temperatures, the
total positive pion density is thus given by
$$
\eqalignno{
n_{\pi^+}=  &n_{\pi^+}(\hbox{thermal}) &\cr
           +& 0.285~n_\eta &\cr
           +&{2\over 3} n_\rho +n_\omega+0.134~n_\phi &\cr
           +&{1\over 3}n_{K^*}+{1\over 3}n_{{\overline K}^*}&\cr
           +&{1\over3}n_\Delta +{1\over 3}n_{\Sigma^*}+{1\over
                              3}n_{\Xi^*}&\cr
           +&{1\over 3}n_{\overline{\Delta}} +
{1\over 3}n_{{\overline \Sigma}^*}+{1\over 3}n_{{\overline \Xi}^*}&\cr
          +&\cdots,           &(7)\cr
}
$$
in which the different factors in front of the resonance densities
are obtained from the corresponding branching ratios into positive
pions. As the temperature is increased still further, more and more terms
will start contributing in the sum in equation (7).
At fixed temperature, the ratio of all pions to thermal pions furthermore
increases rapidly as a function of baryon chemical potential.
This is shown in Fig.4 for two different values of the
temperature $T$. The increase is due to the fact that the number of
baryonic resonances increases with the baryon chemical potential, and
these provide
more pions as decay products. We notice that the fraction
of thermal pions quickly becomes very small;
at high temperatures and densities, it is
negligible compared to the number of pions from
decay products of heavier resonances. It is therefore important to
specify how many such resonances are taken into account in the hadron
gas, since the number of pions keeps increasing as more and
more resonances are included. This dependence has
been investigated and was found to be important
\refss{heinz}{barz}{weber}. In principle
one must include all resonances; in practice, the contributions from
heavy states become important only when the temperature of the system
is relatively high. We show in Fig.3 both the
result obtained from our tables, which include
all observed resonances up to a mass of 2 GeV, and the same ratio
obtained by including only the fundamental hadron multiplets.
 We see that up to temperatures of about 150 MeV, including more
heavy resonances has essentially no effect; at 200 MeV, however,
 the ratio of
all to thermal pions has already increased by about 50\%.
In any case, most pions in the final state of a relativistic heavy
ion collision are certainly decay products of resonances.
The pion spectra therefore do tell us something about equilibrium,
but only in a very indirect way.
\medskip
\noindent{\bf 3.2 Kaons}
\medskip
Kaons are at first sight similar to pions. We see in Fig.5 that
the ratio between the total
number of kaons and those that are solely of thermal origin also rises
rapidly above a temperature of 100 MeV, just as is the case with pions,
even though the rise is not nearly as steep.
 Decay kaons come mainly from
strange mesonic resonances; baryons only rarely have kaons in their
decay products. Even strange baryons do not decay copiously into kaons,
witness the baryons in the decuplet, where only the $\Omega^-$ gives
kaons. As function of the baryon number density,
the ratio of all kaons to thermal kaons therefore does not
behave like the corresponding ratio for pions;
it decreases with increasing $\mu_B$, as seen in Fig.6.
\par
The total density of positive kaons is determined by the relation
$$
n_{K^+}=n_{K^+}(\hbox{thermal}) +0.495~n_{\phi}
+0.5~n_{K^*}+\cdots. \eqno(8)
$$
When the temperature is increased much beyond 150 MeV,
more terms would have to be
added to the right-hand side of the above equation.
As mentioned, the baryons give
almost no contributions, in contrast to the pion case (7), where
they contribute very much. Nevertheless, we
conclude again that a considerable part of the kaons
in the final state of a relativistic heavy ion
collision are not of direct thermal origin, but decay products of
strange mesonic resonances. The situation here is somewhat less
pronounced than it was for pions, however, where those of direct
thermal origin could almost be neglected.
The tables at the end of this paper were calculated using equation (8)
without any further resonance contributions.
\medskip
\noindent{\bf 3.3 Vector Mesons}
\medskip
Vector mesons have attracted much attention, with
the observation of \J{} suppression compared to
the Drell-Yan continuum and of the enhancement of
$\phi$ production compared to that of $\rho$ and $\omega$ \ref{NA38}.
\J{} production lies outside present considerations,
since the \J{} is too heavy to be produced by thermal collisions.
We therefore turn
our attention to the ratio $\phi/(\rho + \omega )$, which is shown in
Fig.7 as a function of temperature. This ratio is independent of the
baryon chemical potential, since non-strange mesons are not
sensitive to the baryonic content of the hadronic gas. Strange mesons are
sensitive to $n_B${} through the strange chemical potential, e.g. via
associated production of $\Lambda$ and $K^+$.
The $\phi$-meson is of particular interest, because its quark structure
never enters the hadronic gas calculations; from a hadronic point of
view, the $\phi$ is just a heavy $\omega$. Only on the
quark level does the $s\bar{s}$-content enter. --
With increasing $\mu_B$, the $\phi/(\rho+\omega)$ ratio
actually drops, because the relative weight of the $\phi$ decreases.
as shown in Fig.8.
\par
The order of magnitude of the ratio $\phi/(\rho+\omega)$ is given by
$$
\phi/(\rho + \omega)\sim {1\over 2}\exp [(m_\rho -m_\phi )/T].
\eqno(9)
$$
For $T = 200$ MeV, this gives a value of $0.14$;
the actual distributions lead to slightly larger values than the
estimate (9).
We have not taken into account vector mesons
originating from the decay of
heavy resonances, since the relevant branching ratios are not very well
established. Their inclusion would lower the ratio, since it will
mainly enhance the number of $\rho$-mesons.
\medskip
\noindent{\bf 3.4 Strange Baryons}
\medskip
Since it is experimentally difficult to separate $\Sigma^0$ and
$\Lambda$, we shall consider these together and denote by $n_{\lambda}$
the overall production rate of the two species.
Again there are non-negligible decay contributions, from the
decuplet baryons $\Sigma^*$ (1.385) and $\Omega^-$ (1.672), which
decay into $\Lambda$'s and $\Sigma$'s. We therefore get
$$
n_{\lambda}=n_{\Lambda}(\hbox{thermal})+{1\over
3}n_{\Sigma}({\rm thermal})+0.9~n_{\Sigma^*}
+
0.68~n_{\Omega}+\cdots,
\eqno(10)
$$
where $n_{\Sigma}$ counts all charge states, so that $n_{\Sigma}/3$
gives us the neutrals. -- In a similar way,
$\Xi$ gets contributions from its decuplet partner the $\Xi^*$ (1.530),
leading to
$$
n_{\Xi}=n_{\Xi}(\hbox{thermal})+n_{\Xi^*}+\cdots.
\eqno(11)
$$
In Fig.9 we show the relative importance of the $\Xi$ (1.32 GeV)
in comparison to the
$\Xi^*$ (1.53 GeV), as a function of temperature for a fixed
$\mu_B$ = 200 MeV. Here we plot
the ratio $(\Xi+\Xi^*)/\Xi$, i.e., (all $\Xi$/thermal $\Xi$).
Again one observes a quick increase of this ratio above temperatures
of 100 MeV, similar to the one observed in Fig.5
for kaons. This shows that even a heavy particle like the $\Xi$ is
strongly influenced by heavier resonances. There thus seems to be no case
where heavy resonances can be neglected. Contrary to the kaon case, the
ratio considered here is independent of the baryon chemical potential,
because both particles are baryons with identical quantum numbers and all
dependence on chemical potentials therefore cancels out. The
tables at the end of this paper were calculated using all
resonances having masses up to about 2 GeV.
\par
To obtain an order of magnitude estimate of
the $\overline{\Lambda}/\Lambda$ ratio, one can use
$$
(\overline{\Lambda}/\Lambda) \sim
\exp [(-2\mu_B+2\mu_S)/T],
\eqno(12)
$$
while for the $\overline{\Xi}/\Xi$ ratio one has
$$
(\overline{\Xi}/\Xi) \sim\exp[(-2\mu_B+4\mu_S)/T].
\eqno(13)
$$
These estimates neglect the dependence on transverse momentum; in this
case, the $\Lambda$ and the $\Lambda^*$ (and similarly the $\Xi$ and the
$\Xi^*$, as well as the corresponding antibaryons) have the
same dependence on $\mu_{B}$. Hence all ``all/thermal" ratios are
here approximately $\mu_{B}$-independent. However, since the momenta
were neglected, these results should be used only as rough
approximations. Nevertheless, the inequality
$$
(\overline{\Xi}/\Xi)>(\overline{\Lambda}/\Lambda)
\eqno(14)
$$
will hold even for the complete calculation, since the ratio (12)
contains a factor $2\mu_S$ less in the exponent than the ratio (13).
\bigskip
\noindent{\bf 4. The Analysis of Equilibrium Behaviour}
\bigskip
We will now illustrate for some specific cases how our tables can be
used to study the degree of equilibration found in heavy ion data, and
what conclusions one might draw from the results.
\medskip
\noindent{\bf 4.1 Strange Baryons}
\medskip
Here we shall consider the data on strange baryons provided by the WA85
\ref{WA85} and NA35 \ref{NA35}
collaborations at CERN. From two measured ratios we can estimate the two
thermal parameters, i.e. the temperature and the baryon chemical
potential. The data of \ref{WA85} provide the
ratios $\overline{\Lambda}/\Lambda = 0.13\pm 0.03$
and $\overline{\Xi}/\Xi =  0.39\pm 0.07$. Each leads to a ``band" in
the $T_f-\mu_B$ plane, which can be determined from our tables. The
result is shown in Fig.10, where we see that the two bands cross in
the rather small shaded region. From this we would get the ranges
170 MeV $\leq T_f \leq$ 220 MeV for the freeze-out temperature and
220 MeV $\leq \mu_B \leq$ 420 MeV for the baryon chemical potential,
if there was equilibrium at the freeze-out of these baryons. Note that
the temperature range is roughly in
accord with temperature estimates from the
transverse momentum distributions, which from large transverse
momentum data give $T_{pt}\simeq 230 - 240$ MeV \ref{WA85}.
 We should further
emphasize that the $p_T$ cut of the experiment does not matter here,
since in both ratios the masses are the same in numerator and
denominator.
\par
So far we have used two data points to estimate two parameters.
To see if these really correspond to the freeze-out parameters of an
equilibrium hadron gas, we need further data. These are provided by
results for $\Xi^-/\Lambda$ and
$\overline{\Xi}$$^-/\overline{\Lambda}$ \ref{WA85}.
Due to the different
masses involved, these ratios have to be corrected for the $p_T$
cuts made in the experiment.
This reduces the measured ratios, since the
$\Xi$-particles have a heavier mass and thus start at higher values
of the transverse mass $m_T=(p_T^2+m^2)^{1/2}$.
We estimate the corresponding correction factor to be
$$
\left[{\Xi^-\over\Lambda}\right]_{\rm corr.}~=~
\left[{\Xi^-\over\Lambda}\right]_{(m_T-{\rm cut})}\times
{\int_{m_{\Xi}}^{\infty}m_Tdm_T\exp (-m_T/T)\over
\int_{m_{\Lambda}}^{\infty}m_Tdm_T\exp (-m_T/T)}
\eqno(15)
$$
For equal masses, this correction factor is unity; this is the case for
the $\overline{\Xi}/\Xi$ and the $\overline{\Lambda}/\Lambda$ ratios.
For
$\overline{\Xi}$$^-/\overline\Lambda$ and $\Xi^-/\Lambda$,
 the correction factor is approximately 0.48.
Using the slopes given by the WA85 collaboration, this leads to
$\Xi^-/\Lambda=0.10 \pm 0.02$ and
$\overline{\Xi}^-/\overline\Lambda=0.29\pm 0.10$
for the corrected ratios.
\par
As above, we now calculate the band in the $T-\mu_B$ plane
corresponding to $\Xi^-/\Lambda=0.10\pm0.02$; in Fig.11 it is combined
with the results shown in Fig.10. We see that this ratio is also
compatible with the $T-\mu_B$ range obtained in Fig.10.
For ${\bar \Xi^-}/{\bar \Lambda}=0.29\pm0.10$, we include in
in Fig.11 only the line corresponding to the value 0.29, without
errors, in order to keep the graphics clear.
It is seen to be compatible with the region of Fig.10 as well.
 We therefore conclude that
the strange baryon ratios as measured by \ref{WA85}
are quite consistent with equilibrium at freeze-out for
$T=200\pm20$ MeV and $\mu_B=320\pm100$ MeV.
Note that this value of $\mu_B$ corresponds to
$n_B$$\simeq 0.3$ fm$^{-3}$, i.e., to about twice nuclear density.
\par
Let us now turn to the ratio $\Lambda/K_s^o= 0.76\pm 0.16$,
 measured by \ref{NA35}.
The resulting line is included in Fig.11;
it is also absolutely consistent with the
previously obtained freeze-out parameters, even though we are now
looking at the ratio of a strange baryon to a strange meson. \par
The picture becomes less consistent, however, if we compare
the $\Lambda$ and
$\overline{\Lambda}$ to all negative hadrons, i.e., to mainly
{\sl non-strange} mesons. The reported measurements \ref{NA35}
$\Lambda/h^-=0.08\pm 0.01$ and $\bar\Lambda/h^- = 0.015\pm
0.005$ lead to the cross-over region shown in Fig.12. Although the
overlap regions in Figs. 11 and 12 just touch, the region in Fig.12 does
point to lower values of the freeze-out temperature as well as baryonic
chemical potential. However, before drawing any conclusions, let us
consider what the ratios among only mesons indicate.
\bigskip
\noindent {\bf 4.2 Strange and Non-Strange Mesons}
\medskip
\refadd{Helios}
\refadd{satz-detlof}\refadd{hungary}\refadd{koch2}
The ratio $\phi/(\rho+\omega)$ was determined in \ref{NA38} and
\ref{Helios}; both get a value around 0.3 - 0.4 at the highest
energy density, and as seen in the dimuon channel. Correcting this
by the dimuon branching ratios for $\phi$ (0.025), $\rho$ and $\omega$
(0.007), we get from \ref{Helios} the value
0.11 $\pm$ 0.01 for the ratio $\phi/(\rho+\omega)$.
 Since all three mesons are non-strange, the thermal ratio
is independent of $\mu_B$. In Fig.7 we had shown the behaviour of the
thermal $\phi/(\rho+\omega)$ as function of \T. The measured value
is included in the figure;
is seen to require very low temperatures, between 100 and 120 MeV.
It is certainly not compatible with the \T$\simeq$200 MeV found
above in the strange baryon analysis.
\par
This conclusion is supported
by an analysis of $K/\pi$ ratio. In \ref{NA35}, the ratio
$K_s^o/\pi^-$, equivalent to
$(K^+ +K^-)/(\pi^+ +\pi^-)$,
is found
to be $0.11 \pm 0.02$ for the integrated data, with a peak value of
$0.15\pm 0.04$ at midrapidity. The resulting bands in the $T_f-\mu_B$
plane are shown in Fig.13; with values between 100 and 120 MeV (for the
integrated data), they also require a considerably lower
temperature than found from strange baryons. The same conclusion had
been reached in an analysis of the BNL data \ref{E802} for the
$K^+/\pi^+$ ratio \refto{satz-detlof}{koch2}, where temperatures
around 100 MeV were needed.
\medskip
\noindent{\bf 4.3 Conclusions}
\medskip
\refadd{phi-heinz1} \refadd{phi-heinz2} \refadd{phi-barz}
\refadd{phi-rafelski}
We have thus found two sets of data, each in itself compatible with
equilibrium hadron ratios, but leading to different freeze-out
parameters. Before we comment on possible interpretations, let us
note some caveats. We have compared experiments covering different
kinematic regions, and we have considered results integrated over
kinematic regions showing quite non-uniform behaviour (mid-rapidity
vs. forward/backward, for example). This is not really meaningful,
and hence everything done here should only serve as an illustration.
Ideally, one should use ratios from symmetric nuclear collisions ($A-A$)
at mid-rapidity and integrated over transverse momenta from zero to
1.5 - 2 GeV. Hopefully, the $Au$-beam at BNL and the $Pb$-beam at CERN
will provide such data. -- The measurements of vector mesons through
dilepton or photon
decays may lead to incorrect rate estimates, if the mesons
are short-lived enough to decay in the medium \refto{phi-heinz1}
{phi-rafelski}.
\par
If we take the result of our analysis serious in spite of these
caveats, they would seem to indicate that the observed strange
baryons freeze-out at an earlier and hence ``hotter and denser"
stage than the mesons. It would be interesting to check if this
is also reflected in different temperatures from transverse
momentum spectra; there seem to be hints in that direction.
\par
In summary: in this paper we have determined and tabulated the
ratios for the various hadron species as function of the freeze-out
temperature \T{} and the baryon number density $n_B$, or equivalently,
the baryonic chemical potential $\mu_B$. These tables should allow a
study of the degree and temporal succession of thermalisation
for the observed hadron species.\par
\bigskip
\noindent {\bf Acknowledgements}
\medskip
One of the authors (J.C.) thanks O. Villalobos Baillie for comments on the
manu\-script. The other one (H.S.) gratefully acknowledges stimulating
discussions with C. P. Singh. \par
\vfill\eject
\bigskip\centerline{\bf References}
\bigskip
\item{\reftag{qm90})}
          {\it ``Quark Matter '90''},
          Proc. of the VIII'th Int. Conf. on
          Ultra-Relativistic Nucleus-Nucleus Collisions, Menton, 1990,
          J.P. Blaizot et al. (editors),
          Nucl. Phys. {\bf A525} (1991).
\smallskip
\item{\reftag{WA85})}
          S. Abatzis et al., WA85 Collaboration,
          Phys. Lett. {\bf B244} (1990) 130
          and Phys. Lett. {\bf B259} (1991) 508.
\smallskip
\item{\reftag{NA35})}
          J. Bartke et al., NA35 Collaboration,
          Z. Phys. {\bf C48} (1990) 191

          R. Stock et al., NA35 Collaboration
          Nucl. Phys. {\bf A525} (1991) 221c.
\smallskip
\item{\reftag{E802})}
          T. Abbott et al., E802 Collaboration,
          Phys. Rev. Lett. {\bf 64} (1990) 847.
\smallskip
\item{\reftag{NA38})}
          C. Baglin etal., NA38 Collaboration,
          Phys. Lett. {\bf B251} (1990) 465 and 472;

          Phys. Lett. {\bf B262} (1991) 362.
\smallskip
\item{\reftag{vesa1})}
          M. Kataja and P.V. Ruuskanen,
          Phys. Lett. {\bf B243} (1990) 181.
\smallskip
\item{\reftag{vesa2})}
          S. Gavin and P.V. Ruuskanen,
          Phys. Lett. {\bf B262} (1991) 326.
\smallskip
\item{\reftag{amis})}
          H.J. Crawford, M.S. Desai and G.L. Shaw,
          Phys. Rev. {\bf D45} (1992) 857.
\smallskip
\item{\reftag{carsten})}
          C. Greiner, P. Koch and H. St\"ocker,
          Phys. Rev. Lett. {\bf 58} (1987) 1825;

          Phys. Rev. {\bf D38} (1988) 2797.
\smallskip
\item{\reftag{strange})}
          {\it ``Strange Quark Matter in Physics and Astrophysics''},

          J. Madsen and P. Haensel (editors),
          Nucl. Phys. {\bf B24} (Proc. Suppl.) (1991).
\smallskip
\item{\reftag{raf})}
          J. Rafelski,
          Phys. Lett. {\bf B262} (1991) 333.
\smallskip
\item{\reftag{heinz})}
          J. Sollfrank, P. Koch and U. Heinz,
          Phys. Lett. {\bf B253} (1991) 256;

          Z. Phys. {\bf C52} (1991) 593;
\smallskip
\item{\reftag{Redlich})}
          J. Cleymans, K. Redlich, H. Satz and E. Suhonen,
          Z. Phys. {\bf C33} (1986) 151.
\smallskip
\item{\reftag{karsch})}
          F. Karsch and H. Satz,
          Phys. Rev. {\bf D22} (1980) 480.
\smallskip
\item{\reftag{Raf-Gat})}
          J. Rafelski,
          University of Arizona preprint AZPH-TH/91-62
\smallskip
\item{\reftag{barz})}
          H.W. Barz, G. Bertsch, D. Kusnezov and H. Schulz,
          Phys. Lett. {\bf B254} (1991) 332.
\smallskip
\item{\reftag{kusnezov})}
          D. Kusnezov and G.F. Bertsch,
          Phys. Rev. {\bf C40} (1989) 2075.
\smallskip
\item{\reftag{brown})}
          G.E. Brown, J. Stachel and G.M. Welke,
          Phys. Lett. {\bf B253} (1991) 19.
\smallskip
\item{\reftag{jsk})}
          J. Schukraft,
          CERN preprint PPE/91-04 (1991) (unpublished).
\smallskip
\item{\reftag{weber})}
          J. Cleymans, E. Suhonen and G.M. Weber,
          Z. Phys. {\bf C53} (1992) 485.
\smallskip
\item{\reftag{Helios})}
          F. Martelli,
          Talk presented at the Lepton-Photon Symposium, Geneva,
          Switzerland, July 25 - August 1, 1991.
\smallskip
\item{\reftag{satz-detlof})}
          J. Cleymans, H. Satz, E. Suhonen and D.W. von Oertzen,
          Phys. Lett. {\bf B 242} (1990) 111.
\smallskip
\item{\reftag{hungary})}
          B. Luk\'acs, J. Zim\'anyi and N.L. Balazs,
          Phys. Lett. {\bf B 183} (1987) 27.
\smallskip
\item{\reftag{koch2})}
          P. Koch as reported by U. Heinz in : {\it ``Phase Structure of
          Strongly Interacting Matter.''}  J. Cleymans (ed.),
          Springer-Verlag, Heidelberg (1990).
\smallskip
\item{\reftag{phi-heinz1})}
          P. Koch, U. Heinz and J. Pi\v{s}ut,
          Phys. Lett. {\bf B243} (1990) 149;

          Z. Phys. {\bf C47} (1990) 477;
\smallskip
\item{\reftag{phi-heinz2})}
          P. Koch and U. Heinz,
          Nucl. Phys. {\bf A525} (1991) 293c.
\smallskip
\item{\reftag{phi-barz})}
          H.W. Barz, B. Friman, J. Knoll and H. Schulz,
          Phys. Lett. {\bf B254} (1991) 315.
\smallskip
\item{\reftag{phi-rafelski})}
          P-Z. Bi and J. Rafelski,
          Phys. Lett. {\bf B262} (1991) 485.
\smallskip
\item{\reftag{satz})}
          J. Cleymans, K. Redlich, H. Satz and E. Suhonen,
          Z. Phys. {\bf C33} (1986) 151.
\smallskip
\item{\reftag{davidson})}
          N.J. Davidson, H.G. Miller, R.M. Quick and J. Cleymans,
          Phys. Lett. {\bf B255} (1991) 105;
          D.W. von Oertzen et al. Phys. Lett. {\bf B274} (1992) 128.
\smallskip
\item{\reftag{koch})}
          P. Koch, B. M\"uller and J. Rafelski,
          Phys. Rep. {\bf 148} (1986) 167.
\smallskip
\vfil\eject
\bigskip\centerline{\bf Figure Captions}\bigskip
\smallskip
\item{Fig. 1:}
Strange chemical
potential $\mu_S$ as a function of baryon chemical potential
$\mu_B$ for temperature values from $T = 150$ MeV to $T=300$ MeV.
The solid line was calculated for a temperature of 10 GeV.
\smallskip
\item{Fig. 2.:} Baryon density $n_B$ as function of baryon chemical
potential $\mu_B$ at temperature $T=200$ MeV for pointlike baryons
(dash-dotted line) and for baryons with hard-core repulsion
(full line).
\smallskip
\item{Fig. 3:} Ratio of all pions to thermal pions as
function of temperature $T$ for baryon chemical potential
$\mu_B = 200$ MeV.
The full line includes the contributions
from the fundamental hadron multiplets only,
the dash-dotted line contributions from all
resonances with masses up to 2 GeV.
\smallskip
\item{Fig. 4:}Ratio of all pions to thermal pions as
function of baryon chemical potential $\mu_B$
for temperatures $T = 170$ MeV (full line) and
$T = 200$ MeV (dash-dotted line); calculations include the fundamental
hadron multiplets only.
\smallskip
\item{Fig. 5:} Ratio of all kaons to thermal kaons as
function of temperature $T$ for baryon chemical potential
$\mu_B = 200$ MeV; calculations include the fundamental hadron
multiplets only.
\smallskip
\item{Fig. 6:}Ratio of all kaons to thermal kaons as
function of baryon chemical potential $\mu_B$ for
temperatures $T = 170$ MeV (full line) and $T=200$ MeV (dash-dotted
line). Calculations include the fundamental hadron multiplets only.
\smallskip
\item{Fig. 7:}Ratio of $\phi$ to $\rho +\omega$ as
function of temperature $T$; calculations include all resonances up to
masses of 2 GeV.
\smallskip
\item{Fig. 8:}Ratio of $\phi$ to $\rho +\omega$ as
function of baryon chemical potential $\mu_B$ at temperature
$T = 200$ MeV, including contributions from all resonances
with masses up to 2 GeV.
\smallskip
\item{Fig. 9:} Ratio of all $\Xi$'s to thermal $\Xi$'s as
function of temperature $T$ for baryon chemical potential
$\mu_B = 200$ MeV; calculations include the fundamental hadron
multiplets only.
\smallskip
\item{Fig. 10:}
$T-\mu_B$ region consistent with the measured ratios for
$\overline{\Lambda}/\Lambda$ and $\overline{\Xi}/\Xi$; the
experimental error limits of the ratios are indicated.
\smallskip
\item{Fig. 11:}
$T-\mu_B$ region consistent with the measured ratios for
$\overline{\Lambda}/\Lambda$, $\overline{\Xi}/\Xi$, $\Xi^-/\Lambda$,
$\overline{\Xi}^-/\overline{\Lambda}$, and $\Lambda/K^o_s$; the
experimental error limits of the ratios are indicated, except for the
last two ratios.
\smallskip
\item{Fig. 12:} $T-\mu_B$ region consistent with the measured ratios
for $\Lambda/h^+$ (dashed line) and for
$\bar\Lambda/h^-$ ratio (full line); the
experimental error limits of the ratios are indicated.
\smallskip
\item{Fig. 13:}$T-\mu_B$ region consistent with the measured ratio for
$K^0_S/h^-$; the experimental error limits of the ratio are indicated.
\vfil\eject
\bye